\documentclass{iau} 

\usepackage{graphicx}
\usepackage{xcolor}
\usepackage[hyphens]{url}
\usepackage{hyperref}
\hypersetup{
     colorlinks  = true,
     linkcolor   = blue, 
     citecolor   = blue,
     filecolor   = blue,
     menucolor   = blue,
     runcolor    = cyan,
     urlcolor    = blue
}
\usepackage[authoryear]{natbib}

\pubyear{2019}
\volume{355} 
\jname{The Realm of the Low-Surface-Brightness Universe}
\editors{D. Valls-Gabaud, I. Trujillo \& S. Okamoto, eds.}
\title[Hydrogen tails, plumes, clouds and filaments] 
      {Hydrogen tails, plumes, clouds and filaments}

\author[B\"arbel S. Koribalski]{B\"arbel S. Koribalski}

\affiliation{CSIRO Astronomy and Space Science, Australia Telescope
    National Facility \\ P.O. Box 76, Epping, NSW 1710, Australia \\ 
    email: {\tt Baerbel.Koribalski@csiro.au}} 

\begin{document}

\maketitle

\begin{abstract}
Here I present a brief review of interacting galaxy systems with extended 
low surface brightness (LSB) hydrogen tails and similar structures. Typically 
found in merging pairs, galaxy groups and clusters, H\,{\sc i} features in 
galaxy surroundings can span many hundreds of kpc, tracing gravitational 
interactions between galaxies and ram pressure forces moving through the 
intra-group/cluster medium. Upcoming large H\,{\sc i} surveys, e.g., with 
the wide-field (FOV = 30 square degrees) Phased Array Feeds on the Australian 
SKA Pathfinder (ASKAP), will provide a census of LSB structures in the Local 
Universe. By recording and comparing the properties (length, shape, H\,{\sc i} 
mass, etc.) of these observed structures and their associated galaxies, we 
can -- using numerical simulations -- try to establish their origin and 
evolutionary path.

%%%%%%%%%%%%%%%%%%%%%%%%%%%%%%%%%%%%%%%%%%%%%%%
%% add here a maximum of 10 keywords, to be taken from the file <Keywords.txt>
%%%%%%%%%%%%%%%%%%%%%%%%%%%%%%%%%%%%%%%%%%%%%%%
\end{abstract}

\firstsection % do not remove

\section{Introduction}

The evolution and transformation of a galaxy is strongly influenced by its
environment, where it is exposed to varying degrees of tidal interactions,
ram pressure stripping, collisions, and ionising radiation. Here I present a 
preliminary review of extended H\,{\sc i} features found in the outskirts of
galaxies and typically observed in galaxy groups and clusters. By gathering 
information on peculiar extensions, such as tidal tails and plumes, bridges 
towards companions, cloud complexes and other debris, we can quantify the 
amount of gas found outside galaxies, investigate the responsible removal 
mechanisms, and determine the age of the encounter. Given the wide range of 
H\,{\sc i} morphologies and kinematics, all seen in projection at one 
snapshot in time, this is not an easy task. Large-scale simulations of 
increasingly complex galaxy evolution scenarios are helpful to explore the 
parameter space. \\

The comprehensive H\,{\sc i} Rogues Gallery, created by Hibbard et al. 
(2001a), contains one of the largest collections of H\,{\sc i} maps and 
classifications of peculiar galaxies. When published it listed 181 systems 
(individual galaxies, pairs, compact groups, etc.) with a total of 400+ 
galaxies. Only a few were added to the "living Gallery" in the following 
years, while further growing the collection and enhancing the visualisation 
of its content would be of great interest. In Koribalski (2004) I highlight 
a few more spiral galaxy systems with H\,{\sc i} bridges, plumes and cloud 
complexes. H\,{\sc i} structures around early-type galaxies are examined 
by Oosterloo et al. (2007) who identify several tails and rings as well as 
large disks and offset clouds. \\

A catalog of long H\,{\sc i} streams ($>$100 kpc; 42 objects), many including 
the host/associated galaxies, is presented by Taylor et al. (2016) who discuss 
the origin of H\,{\sc i} clouds ($<$20 kpc; 51 objects). While this list 
includes many prominent H\,{\sc i} features, some of which are also mentioned 
here, it is dominated by gas-rich galaxy pairs and mergers (e.g., 
NGC~6872/IC\,4970 studied by Horellou \& Koribalski 2007). Detached H\,{\sc i} 
clouds/streams, often without stellar counterparts, are typically found in 
wide-field H\,{\sc i} surveys (e.g., Ryder et al. 2001, Scott et al. 2012)
carried out with single dish telescopes. We are now commencing an era of 
wide-field interferometric H\,{\sc i} surveys with recent results from ASKAP 
and MeerKat, including the discovery of tidal tails, presented by Lee-Waddell 
et al. (2019) and Serra et al. (2019), respectively. \\

While most talks at this wonderful conference focussed on stellar LSB features, 
it is worth highlighting the importance of detailed H\,{\sc i} mapping (incl. 
its kinematics) in complementing our understanding of galaxy transformations 
and evolution. As optical imaging is capable of detecting ever fainter LSB 
structures, we will be able to better analyse the stellar content of H\,{\sc i} 
tidal tails and pinpoints pockets of new star formation as well as, in some 
cases, the formation of tidal dwarf galaxies (TDGs). Interferometric H\,{\sc i} 
imaging can sometimes help to distinguish Galactic cirrus from extragalactic 
features seen in deep optical images. As the number of detected stellar 
streams is growing rapidly, thanks to new instrumentation as well as dedicated 
deep optical imaging of galaxies (e.g., Mart\'inez-Delgado et al. 2008), 
complimentary deep H\,{\sc i} observations are needed.

\section{Overview}

In Table~\ref{koribalski:tab1} I list some of the largest H\,{\sc i} 
structures detected near galaxies in the Local Universe, arranged by their 
approximate size. This is a preliminary collection (40 galaxy systems) --- 
additions by readers are very welcome --- 
with only a limited set of relevant properties. A more comprehensive review
by the author is in preparation. Given the diversity of H\,{\sc i} structures 
observed outside galaxies (tidal tails, rings, plumes, cloud complexes, ...), 
any size/length measurements are approximate and do not take into account 
projection effects. The listed H\,{\sc i} mass is that measured within the 
given size of the structure, which for rings and cloud complexes corresponds 
to their diameter and for filaments, bridges and tidal tails to their length. 
We typically use the same structure designation as given in the cited 
publications. In some case, the structure size includes the associated 
galaxies, where separation of the various components is not easily done. \\

One of the largest H\,{\sc i} structures known in the Local Universe is the 
Magellanic Stream, which together with the Leading Arm spans $\sim$200 
degrees on the sky, tracing the interactions between the Milky Way and the 
Magellanic Clouds. Assuming a distance of 55~kpc, the Magellanic Stream has 
a length of at least 180~kpc and an H\,{\sc i} mass of $2.7 \times 
10^8$~M$_{\odot}$ (Br\"uns et al. 2005), approximately half that of the 
whole Magellanic System. Westmeier \& Koribalski (2008) discover further
wide-spread H\,{\sc i} debris in the vicinity of the stream. 

The most spectacular H\,{\sc i} cloud so far identified in the H\,{\sc i}
Parkes All-Sky Survey is HIPASS J0731--69 (Ryder et al. 2001) with an 
H\,{\sc i} mass of $\sim$10$^9$~M$_{\odot}$ and located $\sim$200~kpc 
north-east of the spiral galaxy NGC~2442; Ryder \& Koribalski (2004) show 
it to consist of numerous clumps along a partial loop. Furthermore, 
Oosterloo et al. (2004) disovered four large H\,{\sc i} clouds scattered 
along a $\sim$500~kpc arc surrounding the elliptical galaxy NGC~1490. 
H\,{\sc i} rings of diameter $\sim$200~kpc have been detected around the 
E/S0 galaxies M\,105 (the `Leo Ring', Schneider et al. 1989, Stierwalt et
al. 2009), NGC~5291 (Malphrus et al. 1997) and NGC~1533 (Ryan-Weber et al.
2004). Their formation mechanism is explored in numerical simulations by 
Bekki et al. (2005a,b) who, inspired by the discovery of HIPASS J0731--69 
(Ryder et al. 2001) find tidal H\,{\sc i} structures resembling rings, 
tails and plumes of sizes up to 500 kpc. 

H\,{\sc i} plumes of size 100 to 200~kpc are found near NGC~4388 (Oosterloo 
\& van Gorkom 2005) and NGC~4254 (M\,99; Kent et al. 2007) in the Virgo 
cluster, east of NGC~3628 in the Leo Triplet (Rots 1978, Stierwalt et al. 
2009) as well as west of NGC~3263 (English et al. 2010). Extended H\,{\sc i} 
tidal tails (length 50 to 100~kpc) are generally associated with merging 
galaxy pairs, e.g., Arp\,85 (M\,51; Rots et al. 1990), Arp\,143 (Appelton 
et al. 1987), Arp\,215 (Smith 1994), Arp\,244, better known as `The Antennae' 
or NGC~4038/9 (Gordon et al. 2001, Hibbard et al. 2001b), and Arp\,270 
(Clemens et al. 1999). A $\sim$100~kpc long tidal arm is detected in the 
spiral galaxy M\,83 (Koribalski et al. 2018), likely formed by accretion
of a dwarf irregular galaxy. Two of the longest H\,{\sc i} tidal tails 
known so far (length 160 and 250~kpc) are both associated with galaxy 
groups in the outskirts of Abell\,1367 (Scott et al. 2012).

For comparison, the largest known galaxy H\,{\sc i} disks are of similar 
size, e.g., Malin\,1 (Pickering et al. 1997), HIPASS J0836--43 (Donley et 
al. 2006) and NGC~765 (Portas et al. 2010). Their H\,{\sc i} diameters 
($>$100~kpc) are in fact as expected from their H\,{\sc i} mass based on 
the H\,{\sc i} size-mass relation (Wang et al. 2016).

\begin{table}
\begin{center}
\caption{Preliminary list of the largest known H\,{\sc i} structures 
         in the Local Universe. }
\label{koribalski:tab1}
% {\scriptsize
\begin{tabular}{lccccl}
\hline 
{\bf Galaxy System} & {\bf H\,{\sc i} structure} & {\bf H\,{\sc i} Size} 
   & {\bf H\,{\sc i} mass} & Distance & References \\ 
             & & [kpc] & [10$^9$ M$_{\odot}$] & [Mpc] \\ 
\hline
 NGC~3226/7 (Arp\,94) & lobes/tails & 730 & 1.5 & 25   & here  \\ 
 ~~~"             & inner tails     & 170 & 0.8 & 25   & Mun95 \\
 NGC~4532/DDO~137 & clumpy tail     & 500 & 0.5 & 16.7 & Koo08 \\ 
 NGC~1490         & clouds/arc      & 500 &  8  & 75   & Oos04 \\
 HCG~44/NGC~3162  & stream/bridge   & 450 & 0.4 & 25   & Ser13 \\ 
 HCG~16/NGC~848   & bridge/envelope & 320 & 13  & 52.9 & Ver01 \\
 FGC~1287 triplet & tidal tail      & 250 & 5.7 & 92   & Sco12 \\
 NGC~4254 (M\,99) & tidal tail      & 250 & 0.8 & 16.7 & Ken07 \\
 IC\,1459         & cloud complex   & 250 &  1  & 29   & Sap18 \\ 
 NGC~2442/2434    & partial ring    & 200 &  1  & 15.5 & Ryd01 \\ 
 NGC~2444/5 (Arp\,143) & tidal tail & 200 & 0.5 & 53   & App87 \\
 M\,105/NGC~3384  & Leo ring        & 200 &  1  & 10   & Sch89 \\
 NGC~3690/IC\,694 (Arp\,299)& narrow tail& 180& 3.3& 48& HY99  \\
 Local Group      & Mag. Stream     & 180 & 0.3 & 0.05 & Br\"u05 \\
 NGC~3263         & Vela cloud      & 175 &  4  & 38   & Eng10 \\
 NGC~5291         & partial ring    & 170 & 50  & 58   & Mal97 \\
 NGC~7252 (Arp\,226) & NW tidal tail& 160 & 2.1 & 63.2 & Hib94 \\
 ~~~"                & E tidal tail & 83  & 1.5 & ~"   & ~~"   \\
 RSCG~42 group    & tidal tail      & 160 & 9.3 & 92   & Sco12 \\
 HCG~92 (Arp\,319)& tidal tail/arc  & 125 & 6.5 & 85   & Wil02 \\
 Cartwheel Galaxy & plume/bridge    & 110 &  3  & 124  & Hig96 \\
 NGC~4388 (M\,86) & complex tail    & 110 & 0.3 & 16.7 & OG05 \\
 M\,81 group      & tidal bridges   & 100 & 0.7 & 3.6  & Yun94 \\
 NGC~2146         & streams + debris& 100 & 4.6 & 12.2 & Tar01 \\
 HI\,1225+01      & bridge/tail     & 100 & 2   & 20   & Che95 \\
 NGC~6215/6221    & bridge          & 100 & 0.3 & 18   & KD04 \\
 NGC~3256         & tidal tails     & 100 &  5  & 38   & Eng03 \\
 NGC~7582         & tidal tails     & 100 &  2  & 20   & Kor96 \\
 M\,51 (Arp\,85 ) & tidal tail      & 90  & 0.5 & 9.6  & Rot90 \\
 NGC~4038/9 (Arp\,244)&southern tail& 90  & 2.5 & 19.2 & Gor01 \\
 ~~~"             & northern tail   & 30  & 0.2 & ~"   & ~~"   \\
 NGC~4111         & plume           & 72  & 0.1 & 15   & VZ01  \\
 Leo Triplet (Arp\,317) & tidal tail& 70  & 0.5 & 6.7  & Rot78 \\
 NGC~1316 (Fornax\,A) & tidal tails & 70  & 0.7 & 20   & Ser19 \\
 CGCG097--087     & tidal tail      & 70  &     & 92   & Sco12 \\
 NGC~3395/6 (Arp\,270) & tidal tail & 63  & 0.5 & 21.7 & Cle99 \\
 NGC~4424         & narrow tail     & 60  &0.04 & 15.2 & Sor17 \\
 HCG~31           & southern tail   & 56  & 6.4 & 54.3 & Ver05 \\
 ~~~"             & NW tail         & 50  & 4.0 & ~"   & ~~"   \\
 ~~~"             & NE tail         & 35  & 0.3 & ~"   & ~~"   \\
 NGC~2782         & plume           & 54  & 1.4 & 34   & Smi94 \\
 NGC~3310 (Arp\,217) & southern tail& 51  &0.27 & 13   & KS01  \\
 ~~~"             & northern tail   & 23  &0.23 & ~"   & ~~"   \\
 NGC~4694         & tail/bridge     & 50  & 0.9 & 17   & Duc07 \\
 NGC~4026         & plume/filament  & 38  & 0.2 & 17.1 & VZ01  \\
 IC\,2554         & tidal plume     & 30  & 0.7 & 16   & KGJ03 \\
\hline 
\end{tabular}
%  }
\end{center}
%\vspace{1mm}
% \scriptsize{
% {\it Notes:}\\
%  $^1$...
\end{table}

\begin{figure*} % Figure 1
\begin{tabular}{cc}
  \includegraphics[scale=0.32]{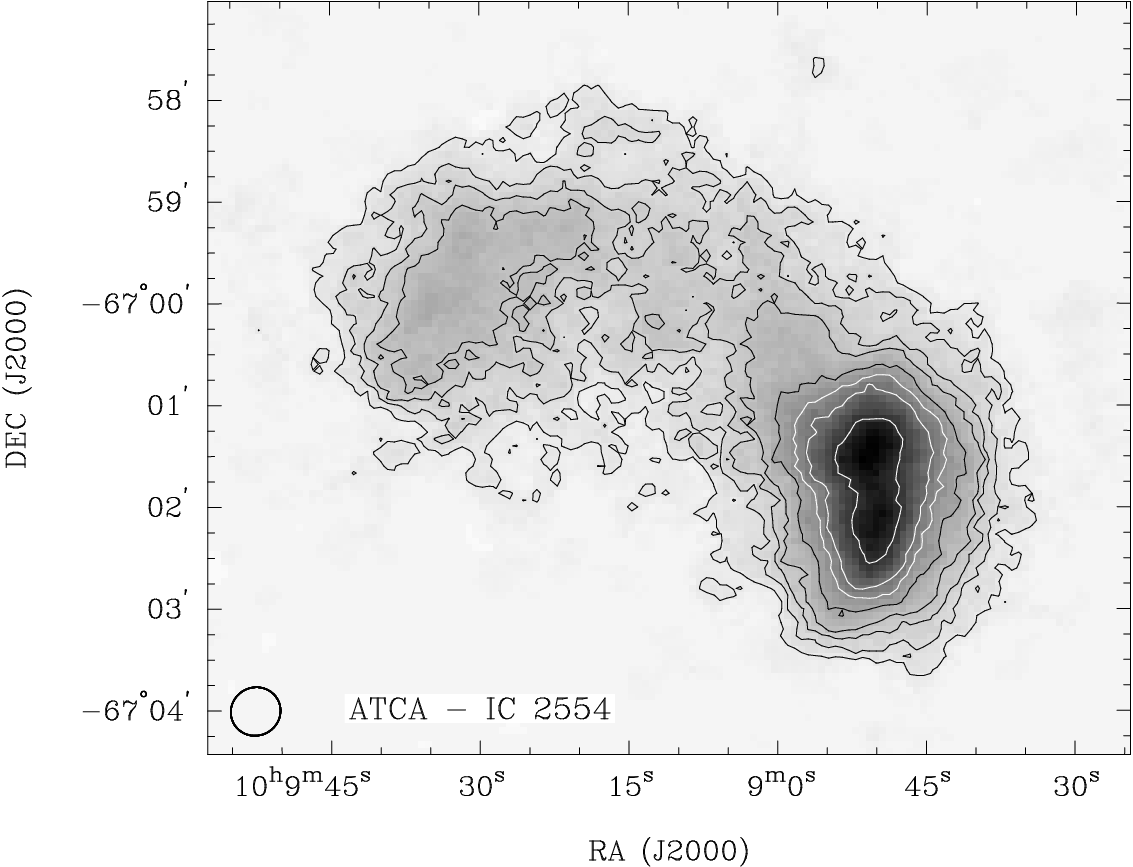} &
  \includegraphics[scale=0.27]{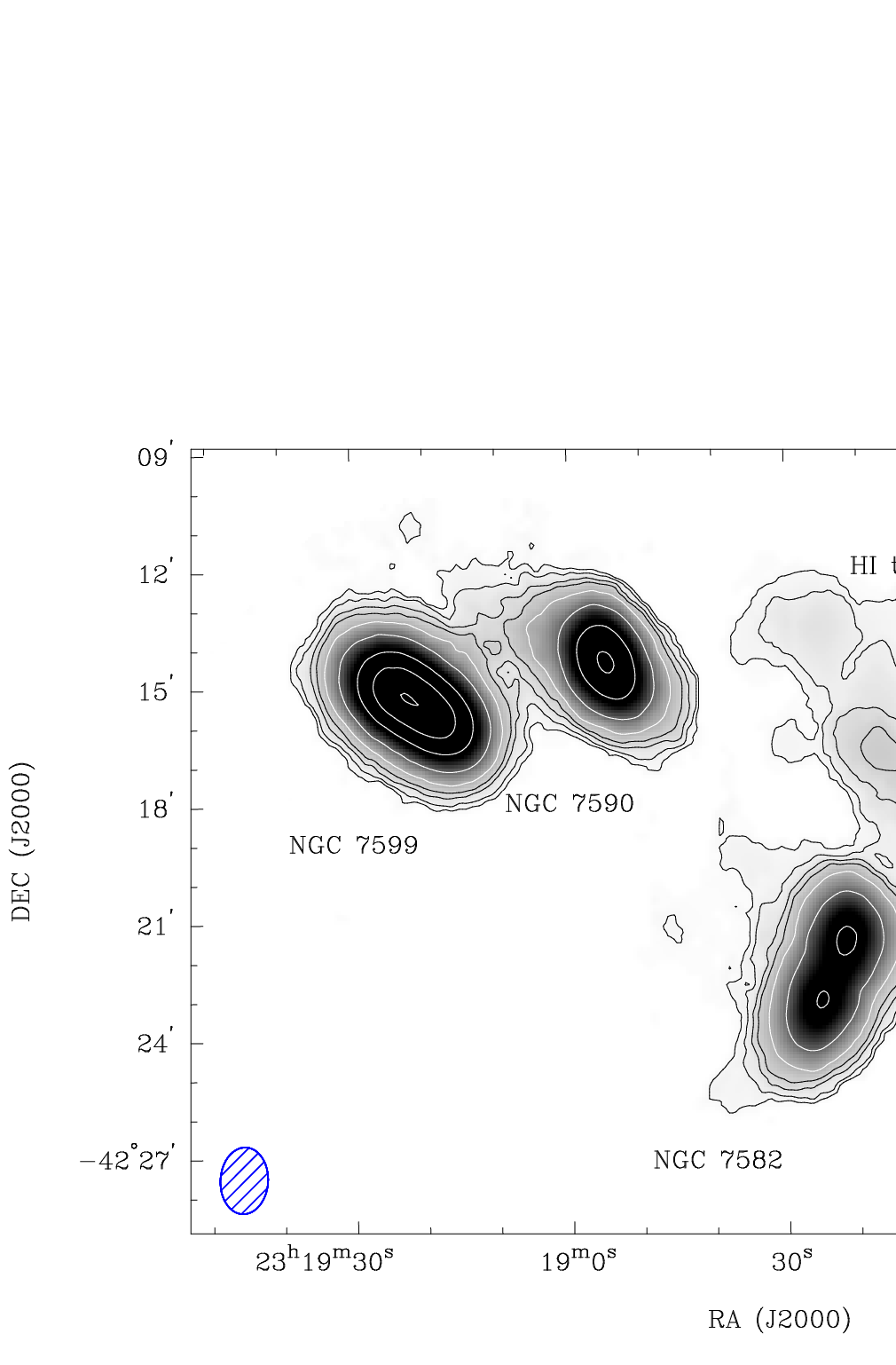} \\
\end{tabular}
\caption{ATCA H\,{\sc i} intensity maps of the IC\,2554 galaxy system 
   (HIPASS J1008--67; left) and the NGC~7582 triplet (HIPASS J2318--42; 
   right). The prominent H\,{\sc i} plume east of IC\,2554 has an 
   H\,{\sc i} mass of close to 10$^9$~M$_{\odot}$ (Koribalski, Gordon 
   \& Jones 2003), while the broad tidal tails north of NGC~7582 
   contain nearly $2 \times 10^9$~M$_{\odot}$ (Koribalski 1996).}
\end{figure*}

\subsection{HI streams and debris}

\begin{itemize}

\item {\bf HCG~16/NGC~848}. --- Verdes-Montenegro et al. (2001) find extended 
H\,{\sc i} emission (stretching $\sim$320 kpc from NW to SE) in HCG~16 
(Arp\,318; HIPASS J0209--10) and surroundings, enveloping the gas-rich group 
members NGC~833/5, NGC~838/9 and NGC~848. They estimate a total group mass of 
at least $2.6 \times 10^{10}$~M$_{\odot}$ and estimate that $\sim$50\% of the 
gas is now dispersed in tidal features, including a narrow H\,{\sc i} bridge
from NGC~833/5 to NGC~848. \\

\item {\bf HCG~44/NGC~3162}. --- Serra et al. (2013) find H\,{\sc i} debris 
scattered throughout HCG~44 (HIPASS J1017+21) and its environment, some of it 
forming a stream of at least 200~kpc length. An even larger H\,{\sc i} bridge
of length $\sim$450~kpc is detected in improved HIPASS data of the area (see 
Fig.~2), extending between HCG~44 and the asymmtric spiral galaxy NGC~3162 
(HIPASS J1013+22). We measure an H\,{\sc i} mass of at least $4 \times 10^8$
M$_{\odot}$. See Leisman et al. (2016) for an in-depth study of the Leo Group
using Arecibo H\,{\sc i} multibeam data. \\

\item {\bf NGC~2146}. --- Taramopulos et al. (2001) detect extended 
H\,{\sc i} streams to the north and south of the peculiar spiral NGC~2146,
possibly formed by a tidal interaction with an unseen H\,{\sc i}-rich LSB 
companion $\sim$800~Myr ago.
They derive an H\,{\sc i} mass of $6.2 \times 10^9$~M$_{\odot}$ for the 
system, half of which resides in the southern stream. The northern stream
and the galaxy contain approximately one quarter of the mass each. \\

\item {\bf IC\,1459}. --- Saponara et al. (2018), using the Australia 
Telescope Compact Array (ATCA), detect a large H\,{\sc i} cloud complex 
near the elliptical galaxy IC\,1459 (HIPASS J2257--36) as well as further 
H\,{\sc i} debris in the wider surroundings. The total extent of these 
tidal debris, extending from the north-east of the spiral galaxy NGC~7418A 
(HIPASS J2256--36a) to the north-east of IC\,1459 is at least $\sim$250~kpc 
(M$_{\rm HI} \sim 10^9$~M$_{\odot}$). 

The IC\,1459 galaxy group was our first WALLABY science target for H\,{\sc i}
observations with an early array of only six PAF-equipped ASKAP antennas.
Serra et al. (2015b) report the detection of 11 spiral galaxies and three 
H\,{\sc i} clouds, two near the edge-on galaxy IC\,5270 (HIPASS J2258--35)
and one near the spiral galaxy NGC~7418 (HIPASS J2256--37). At the time we 
were only observing with nine of the 
36 ASKAP beams (each with FWHM $\sim$1 degr). Comparison of ASKAP and 
HIPASS data showed that the two ASKAP discovered compact clouds (H\,{\sc i} 
mass $\sim$ 10$^9$~M$_{\odot}$ each) are embedded within a larger H\,{\sc i} 
plume containing an additional $\sim$10$^9$~M$_{\odot}$ of diffuse 
H\,{\sc i} emission.  While the low angular resolution of HIPASS can make 
it difficult to identify H\,{\sc i} tails, streams or plumes very close to 
H\,{\sc i}-rich galaxies, an offset between the optical and the H\,{\sc i} 
distributions is a promising signature.  \\

\item {\bf NGC~3263}. --- English et al. (2010), also using the ATCA, find 
a large H\,{\sc i} cloud complex ("the Vela Cloud") with a mass of $3-5 
\times 10^9$~M$_{\odot}$ and size of 100 kpc $\times$ 175 kpc. The cloud 
resides south of the merging galaxy NGC~3256 (HIPASS 1027--43), studied in 
detail by English et al. (2003), and west of the tidally disturbed galaxy 
NGC~3263 (HIPASS J1029--44b).

\end{itemize}

\subsection{HI rings and partial rings}

\begin{itemize}

\item {\bf NGC~1490}. --- Oosterloo et al. (2004) report the discovery of 
several large H\,{\sc i} clouds along a $\sim$500~kpc long arc around the 
elliptical galaxy NGC~1490. The total H\,{\sc i} mass of the clouds, as 
determined from Parkes and ATCA data, is $\sim$8 $\times 10^9$~M$_{\odot}$. 
Deep optical images reveal a very low surface brightness optical counterpart 
in the core of the largest H\,{\sc i} cloud, suggesting one of the highest 
known H\,{\sc i} gas to light ratios. Oosterloo et al. favour a tidal origin 
for this arc-shaped cloud ensemble (HIPASS J0352--66), possibly similar to 
that of the Leo Ring (see Bekki et al. 2005a,b). \\

\item {\bf NGC~2434 group}. --- Ryder et al. (2001) discovered one of the 
most massive star-less (= "dark") H\,{\sc i} clouds, named HIPASS J0731--69 
(see also Koribalski et al. 2004). It has an H\,{\sc i} mass of nearly 
10$^9$~M$_{\odot}$ and is located in the NGC~2434 galaxy group, roughly 
between the one-armed spiral NGC~2442 (HIPASS J0736--69) and the galaxy 
pair NGC~2397/2397B (HIPASS J0721--69). Its closest neighbour is the gas-poor 
elliptical galaxy NGC~2434, located 23$'$ ($\sim$100~kpc) south-east 
of HIPASS J0731--69. Closer inspection of the HIPASS data cube
reveals H\,{\sc i} emission along a partial ring, spanning about $40'
\times 20'$ (ie. 200 kpc $\times$ 100 kpc), connecting to NGC~2442. 
No stellar counterparts have been detected towards any part of this extended 
H\,{\sc i} structure. Only partial GALEX coverage is available for
this area and does not reveal any associated ultra-violet emission.
See Bekki et al. (2005a,b) for possible formation scenarios. \\

\item {\bf NGC~5291}. --- Malphrus et al. (1997) discover a 170~kpc (N-S)
$\times$ 100~kpc (E-W) diameter, partical H\,{\sc i} ring associated with 
the peculiar early-type galaxy NGC~5291 and its companion, the Seashell 
galaxy. Bournaud et al. (2007), using numerical simulations, suggest the 
H\,{\sc i} feature was created by a head-on collision with a massive 
elliptical galaxy $\sim$360~Myr ago. The NGC~5291 system is also detected 
in HIPASS (Koribalski et al. 2004) and catalogued as HIPASS J1347--30.

\end{itemize}

\subsection{Two HI tails}

\begin{itemize}

\item {\bf NGC~3226/7 (Arp\,94)}. --- The merging galaxy system NGC~3226/7 
appears to have two sets of tidal tails, possibly one old and one young pair, 
offset by $\sim$45 degr on the sky (see Fig.~2). Discovered in HIPASS data,
while investigating the nearby HCG\,44/NGC~3162 system (Serra et al. 2013),
these giant H\,{\sc i} tails/lobes span $\sim$100$'$ (ie. 730~kpc for a 
distance of 25~Mpc). The whole system contains at least $1.5 \times 10^9$ .
M$_{\odot}$, at least half of which resides in the north-eastern lobe. The
latter was first discovered in northern HIPASS (Wong et al. 2006) and 
catalogued as HIPASS J1025+20.
The giant H\,{\sc i} lobes are likely quite old ($>$1.5 Gyr) and probably 
falling back towards the merger core as shown in disk-disk merger simulations 
by Di Matteo et al. (2005). The inner H\,{\sc i} tidal tails, discovered by 
Mundell et al. (1995), span only $\sim$20$'$ (oriented north to south) and are 
likely relatively young, recently formed and still fast moving. Deep optical 
images of this area reveal a very faint stellar counterpart to the northern 
H\,{\sc i} tail (e.g., Appleton et al. 2014; Duc et al. 2015). Similarly deep 
images are not yet available for the outer H\,{\sc i} tails/lobes; neither DSS 
nor SDSS images show any likely optical counterparts. \\

\begin{figure} % Figure 2
\begin{center}
  \includegraphics[width=1.2\textwidth]{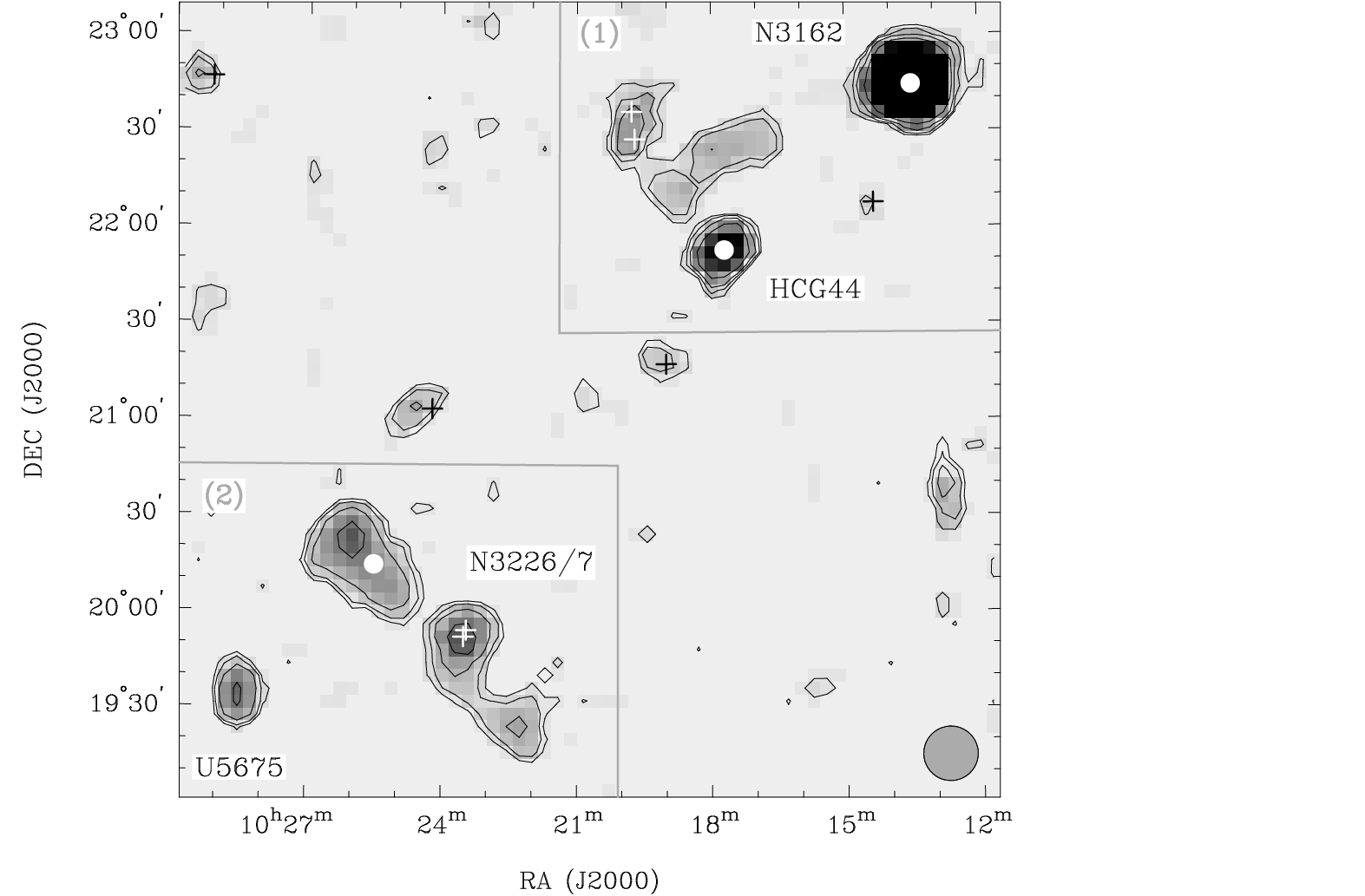} 
\caption{HIPASS intensity map of (1) the long H\,{\sc i} stream reaching 
  from HCG~44 (HIPASS J1017+21) towards the peculiar spiral galaxy NGC~3162 
  (HIPASS J1013+22) and (2) the giant bipolar lobes associated with the 
  merging galaxy pair NGC~3226/7 (Arp\,94). The full NE-–SW extent of the 
  giant H\,{\sc i} lobes is $\sim$100$'$ (ie., 730~kpc); its southern 
  component (HIPASS J1025+20) was first detected by Wong et al. (2006). 
  Hints of the 450~kpc-long H\,{\sc i} stream between HCG~44 and NGC3162
  were first detected by Serra et al. (2013). --- The HIPASS contours are 
  0.5, 1, 2 and 4 Jy/beam km/s. Here I use a convolved Parkes beam of 
  17$'$, displayed in the bottom right corner.}
\label{koribalski:figX}
\end{center}
\end{figure}

\item {\bf NGC~7252 (Arp\,226)}. --- Hibbard et al. (1994) detect two 
extended, countermoving H\,{\sc i} tidal tails in the "Atoms-for-Peace" 
galaxy NGC~7252 with a combined length $\sim$243 kpc, while no H\,{\sc i} 
is found in the central stellar body. While the stars and H\,{\sc i} 
emission in the eastern (E) tail have similar extent, the H\,{\sc i} 
emission in the north-western (NW) tail extends $\sim$50\% further than 
the stars. This is one of many signatures of the original spiral galaxy 
progenitors whose major merger resulted in an elliptical galaxy remnant, 
notable for its stellar shells and loops (see also Hibbard \& Mihos 1995). \\

\item {\bf NGC~7582}. --- Koribalski (1996) find extended H\,{\sc i} tails 
associated with the starburst spiral galaxy NGC~7582, which is a member of the 
Grus Quartet. The two tails, probably stripped from NGC~7582's outer disk by 
tidal interactions, span $\sim$18$'$ (105 kpc). Their kinematical signature 
suggests a connection to the southern (approaching) side of NGC~7582. 
Together the two tidal tails have an H\,{\sc i} mass of nearly $2 \times 
10^9$~M$_{\odot}$, which is approximately 15\% of the total H\,{\sc i} mass 
detected in the Grus Quartet. The latter consists of four large spiral galaxies: 
a triplet consisting of NGC~7582, NGC~7590, NGC7599 (HIPASS J2318--42; see 
Fig.~1) and the face-on starburst galaxy NGC~7552 (HIPASS J2316--42) to the 
west (not shown here). H\,{\sc i} is also detected in a dwarf irregular 
companion galaxy (marked Dwarf\,1) and located just north of NGC~7582.

\end{itemize}

\subsection{One-sided HI tails and plumes}

\begin{itemize}
\item {\bf NGC~2782}. --- Smith (1994) finds a 54 kpc long 
H\,{\sc i} plume on the western side of the peculiar spiral galaxy NGC~2782, 
opposite to its eastern stellar tail. Using kinematic modelling she 
demonstrates that an off-center head-on collision between two galaxies of 
mass ratio $\sim$4:1 about 200~Myr ago can reproduce the observed system. In 
such a scenario, the smaller companion becomes the eastern extension and the 
long western plume consists of gas pulled out from the larger galaxy.
Smith (1994) derives an H\,{\sc i} mass of $1.4 \times 10^9$~M$_{\odot}$ for
the western plume, which is $\sim$40\% of the system's total H\,{\sc i} mass.
Knierman et al. (2013) find young star clusters forming within both tidal 
tails and compare their formation with those in the tidal arms/tails of other 
mergers, e.g. NGC~6872/IC\,4970 (Horellou \& Koribalski 2007) and NGC~3256 
(English et al. 2003). Related to this is the study of tidal dwarf galaxy 
formation by Lelli et al. (2016) in the NGC~4694, NGC~5291 and NGC~7252 
systems. \\

\item {\bf NGC\,3690/IC\,694 (Arp\,299)}. --- Hibbard \& Yun (1999) find a
180~kpc long northern H\,{\sc i} tail/plume in Arp\,299, likely formed via
a merger of two spiral galaxies $\sim$750~Myr ago. While this tidal tail is 
barely visible in the optical, it is accompanied by a remarkable LSB stellar
tail, $\sim$20~kpc offset, with very little H\,{\sc i} emission. \\

\item {\bf IC\,2554}. --- Koribalski, Gordon \& Jones (2003) detect a large 
H\,{\sc i} cloud between the peculiar galaxy IC\,2554 and the elliptical 
galaxy NGC~3136B. The cloud is connected to the northern (approaching) end
of IC\,2554 and forms an arc-shaped plume to the east (see Fig.~1). The 
IC\,2554 system (HIPASS J1008--67) contains a total H\,{\sc i} mass of $2 
\times 10^9$~M$_{\odot}$, a third of which resides in the cloud. The role 
of NGC~3136B, situated just east of the 30~kpc sized H\,{\sc i} cloud and 
undetected in H\,{\sc i}, remains unclear as its systemic velocity is 
$\sim$500 km/s higher than that of IC\,2554, suggesting it is not a likely 
interaction partner. Alternately, IC\,2554 may be a merger remnant. 

\end{itemize} 

\section{Summary and Outlook}
The large number and diversity of H\,{\sc i} structures observed near
galaxies within groups and clusters suggests that such features are common. 
A review of currently known H\,{\sc i} structures, many without detected
stellar counterparts, is timely in view of upcoming large-scale H\,{\sc i}
surveys, e.g. WALLABY ($\delta < +30^{\rm o}$; $z < 0.26$; resolution 
$\sim$30$''$ and 4 km/s) on ASKAP (Koribalski 2012), which will detect a 
wealth of such extended LSB H\,{\sc i} structures. It will assist in our 
preparations for the exploration and analysis of very large 3D data volumes, 
e.g., searching for large H\,{\sc i} structures using our modular Source 
Finding Application (SoFiA; Serra et al. 2015a) as well as new AI/ML 
algorithms (e.g., Gheller et al. 2018). We also like to be able to recognise 
and classify peculiar objects from the shape of their integrated H\,{\sc i} 
spectra alone, as most will remain spatially poorly resolved. Simulated deep 
H\,{\sc i} images, like those of Arp\,299 at redshifts $z = 0.1$ to 0.6 
(Hibbard 2000), will help to assess the detectability of H\,{\sc i} tails 
and other features at large distances. \\

The MeerKat H\,{\sc i} survey of the Fornax cluster (led by Paolo Serra)
will have excellent angular and velocity resolution as well as sensitivity 
to shed new light on the outskirts of galaxy disks impacted by ram pressure 
and tidal forces. It will also deliver a large number of "dark" H\,{\sc i} 
clouds such as the debris recently found in the vicinity of NGC~1316 
(Fornax\,A) by Serra et al. (2019). Some of these H\,{\sc i} clouds are 
also detected in the improved HIPASS data cubes. In the Local Universe, the
combination of single-dish and interferometric data will likely deliver 
the most reliable H\,{\sc i} measurements. Where initially only the densest 
H\,{\sc i} clumps are detected and detectable with an interferometer, 
single dish telescopes can also detect the diffuse H\,{\sc i} emission,
often spread over much larger scales. Notable are the three H\,{\sc i}
sources in the vicinity of Cen\,A (Koribalski et al. 2018; their Fig.~14),
whose nature and origin requires further investigation. \\

Recent ASKAP 21-cm observations of the Eridanus galaxy group/cluster with 
the full array of PAF-equipped telescopes ($36 \times$ 12-m antennas, 
36 beams) are delivering a large number of new galaxy detections and 
hopefully also some new H\,{\sc i} clouds, streams and tidal tails.

\newpage 

\begin{discussion}

\discuss{Sarah Pearson}{Deep optical images of the galaxy M\,51 show 
  faint stellar extensions, e.g. to the north and north-west --- see 
  talk by Chris Mihos --- very different to the location of its 90-kpc
  long H\,{\sc i} tail. How do you explain this\,?} 
  Such differences are important signatures of the disk composition 
  and size of the progenitor galaxies as well as the physical processes 
  involved in their interaction/merger. Furthermore, it is likely that
  multiple merging events took place, resulting in a rich set of loops,
  shells and tidal tails. Prominent offsets between H\,{\sc i} and
  stellar features are also observed in, e.g., Arp\,299 (Hibbard \& Yun 
  1999) and NGC~2782 (Smith 1994). \\

\discuss{Noah Brosch}{What can you say about the gas kinematics of 
  the tidal features you showed during your talk\,?} 
  Analysing the H\,{\sc i} kinematics of tidal tails, plumes, streams, 
  etc. is essential to understanding the galaxy interactions by which 
  they formed. The velocity field often reveals the 3D shape of these
  features as well as their continuity or lack thereof. In particular, 
  the observed kinematic signatures allow to distinguish between 
  different simulations that may have resultied in similar morphologies. 
  In most cases, medium to high velocity resolution ($\sim$4--20 km/s)
  H\,{\sc i} observations are available for the listed galaxy systems. 
  WALLABY -- the ASKAP H\,{\sc i} All Sky Survey -- has a resolution 
  of $\sim$30$''$ and 4 km/s (Koribalski 2012; Lee-Waddell et al. 2019). \\

\discuss{Ignacio Trujillo}{Deep optical images of the LSB galaxy 
  Malin\,1 show a very narrow stellar feature towards the north-east, 
  reaching just beyond the spiral pattern; is any H\,{\sc i} emission 
  detected in that region\,?} 
  The known H\,{\sc i} disk of Malin\,1 (see Pickering et al. 1997) 
  does not extend to the tip of this LSB feature, neither is any
  peculiar H\,{\sc i} emission detected in that direction. \\

\discuss{John Beckman}{Are the ASKAP Phased Array Feeds cooled ? 
  What is the T$_{\rm sys}$\,?} 
  All 36 ASKAP PAFs are passively cooled using solid-state thermo-electric 
  Peltier modules. The effective system temperature (T$_{\rm sys}/\eta$)
  is $\sim$70--75~K across the central part of the band (1.0 -- 1.4~GHz)
  and somewhat higher near the edges of the full band (0.7 -- 1.8~GHz). 
  A bandwidth of $\sim$300~MHz is available for each observation, divided
  into $\sim$17\,000 channels, resulting in 4~km/s velocity resolution. 
  The ASKAP data rate is huge and, at least initially, it will not be 
  possible to keep all spectral line visibilities. \\

\discuss{Samuel Boisssier}{What can you say about associated GALEX 
  ultraviolet emission\,?}
  During my studies of the outer H\,{\sc i} disks of nearby galaxies
  I have found GALEX FUV and NUV imaging of great value, allowing the
  detection of star forming regions often well beyond those visible in 
  optical sky surveys. The XUV-disks of spiral galaxies are typically 
  accompanied by 2X-H\,{\sc i} disks, where dense H\,{\sc i} clumps 
  in the galaxy outskirts are nearly always detected in GALEX FUV 
  emission. The same appears to be the case for H\,{\sc i} tail tidal, 
  which often shown GALEX emission associated with the densest H\,{\sc i} 
  clumps. \\

\end{discussion}

\end{document}